\documentclass[12pt,preprint]{aastex}

\begin{document}

\title{CIRCUMSTELLAR Na\,I AND Ca\,II LINES OF TYPE Ia SUPERNOVAE
IN SYMBIOTIC SCENARIO}

\author{N. N. Chugai}
\affil{Institute of Astronomy, RAS, Pyatnitskaya 48, 119017 Moscow, Russia}
\email{nchugai@inasan.ru}

\begin{abstract}

Formation of circumstellar resonant lines of Na\,I and Ca\,II in 
type Ia supernovae is studied for the case, when supernova explodes 
in a binary system with a red giant. The model suggests 
a spherically-symmetric wind and takes into account ionization and 
heating of the wind by X-rays from the shock wave and by gamma-quanta 
of $^{56}$Ni radioactive decay. For the wind density typical of the 
red giant the expected optical depth of the wind in Na\,I lines 
turnes out low ($\tau<10^{-3}$) to detect the absorption.
For the same wind densities the predicted optical depth of 
Ca\,II 3934 \AA\ is sufficient for the detection ($\tau>0.1$). 
I conclude that the absorption lines detected in SN~2006X 
cannot form in the red giant wind; they are rather related 
to clouds at distances larger than the dust evaporation radius 
($r>10^{17}$ cm). From the absence in SN~2006X of Ca\,II absorption lines 
not related with the similar Na\,I components I derive the 
upper limit of the mass loss rate by the wind with velocity 
$u$: $\dot{M}<10^{-8}(u/10\,\mbox{km s}^{-1})~M_{\odot}$ yr$^{-1}$.

\end{abstract}

\newpage

\section{INTRODUCTION}

Thermonuclear supernovae (SN~Ia)  are the result of explosion of 
 carbon-oxygen white dwarf that attains the Chandrasekhar mass of 
$1.4~M_{\odot}$ due to 
mass accretion in a binary system (Wheelan \& Iben 1973). 
The question, which types of binaries are responsible for SN~Ia, remains 
a subject of debates. A supernova explosion 
in a symbiotic system ("symbiotic scenario") in which a donor is a 
red giant is among possible options 
(Tutukov \& Yungelson 1976; Iben \& Tutukov 1984; Hachisu \& Kato 2000).
 This scenario is a subject of 
a special interest because of the possibility of observational tests.

One of the outcomes of the symbiotic scenario might be the presence 
of the red giant matter in the SN~Ia envelope  (Chugai 1986; 
Livne et al. 1992; Marietta et al. 2000) which could be observed 
as a narrow H$\alpha$ emission at the nebular stage. Until now 
this line has not been detected. 
More obvious consequence of the symbiotic scenario would be the presence 
of a slow red giant wind. The wind could be revealed through 
the radio and X-ray emission originated from the supernova expansion 
in the circumstellar (CS) gas.
The search in radio (Panagia et al. 2006) and X-ray (Hughes et al. 2007) 
bands provides as yet only an upper limit of fluxes.
The signature of the red giant wind could be present at the early stage of 
supernova in the optical as the narrow H$\alpha$ emission, 
which however also has not been found as yet (Cumming et al. 1996). 
Finally, the red giant wind 
could be in principle observed as variable CS  Na\,I and Ca\,II absorption 
lines with 
radial velocities of several tens of km s$^{-1}$ against the SN~Ia 
quasicontinuum.

Recently, Patat et al. (2007) detected variable Na\,I 5890, 5896 \AA\ 
absorption lines with low expansion velocities in the 
spectrum of SN~2006X, a normal type Ia supernova. These absorptions 
being identified with a red giant wind compose a serious argument in favour 
of the symbiotic scenario at least in the particular case of SN~2006X
(Patat et al. 2007). Given both the importance of this result and the 
fact that the formation of the CS absorption lines of 
Na\,I and Ca\,II in the red giant wind after SN~Ia explosion
has not been studied until now, it would be of interest to 
understand what is the expected intensity of these lines in the 
symbiotic scenario and whether these line could be observable. 
 
Here I propose a simple model to describe physical conditions in the 
red giant wind after the SN~Ia explosion, which eventually 
would permit us to calculate 
intensities of CS absorption lines of Na\,I and Ca\,II against the supernova 
background. I start with the analysis of physical conditions in 
the preexplosion wind (section 2.1). I then calculate temperature and 
ionization of the wind after the supernova explosion using time-dependent 
model and taking into account absorption of X-rays from shock wave 
and Compton scattering of gamma-quanta from radioactive $^{56}$Ni decay 
(section 2.2). This provides us a possibility to find the ionization 
fraction of Na\,I and Ca\,II and the optical depth in resonant lines of these
ions as a function of time (section 2.3). In the last 
section I discuss the observational test for the symbiotic scenario 
of SN~Ia and compare results with observations of these lines in SN~2006X.

\section{MODEL AND RESULTS}

Modeling of circumstellar Na\,I and Ca\,II lines in SN~Ia spectra 
in a context of the symbiotic scenario suggests generally 
a solution of two 
problems: (1) determination of an initial state of the CS gas 
before the supernova explosion and (2) calculation of  
conditions in the wind after the supernova explosion. In fact, as we 
will see below, the first issue is of little significance for the final result.
Nevertheles we will consider this problem using simple models of the 
wind and the central source of ionizing radiation in the symbiotic 
system. The second problem is by far more important for the final result 
and its solution requires the consideration of all the essential physics of 
the wind ionization and heating as well as of the ionization of 
Na\,I and Ca\,II.

\subsection{Wind before supernova explosion}

Red giant wind properties in the symbiotic scenario of SN~Ia 
are poorely known; even for the well studied galactic symbiotic binaries 
the mass-loss rate is determined with a large uncertainty. 
The mass-loss rate estimates lie in the range of
$10^{-8}-10^{-6}~M_{\odot}$ yr$^{-1}$ (Korreck et al. 2007), while 
the wind velocities are in the range of 10--50 km s$^{-1}$. We adopt 
below the wind velocity $u=30$ km s$^{-1}$.
An orbital motion and and a fast bipolar wind from the disk 
can result in asphericity of the slow wind outside the binary orbit.
We, however, consider a stationary spherical wind with the 
density $\rho=w/4\pi r^2$. Results obtained for this model could be used 
also to estimate conditions in the CS gas with a more 
complex structure. It is convenient to deal  with 
the parameter $\omega$ defined by the relation 
$w=6.3\times10^{13}\omega$ g cm$^{-1}$. The mass loss rate in 
terms of $\omega$ is $\dot{M}=10^{-6}\omega u_{10}~M_{\odot}$ yr$^{-1}$, 
where $u_{10}$ is the wind veloctity in units of 10 km s$^{-1}$.

Before the supernova explosion the hydrogen in the wind is 
ionized by the radiation of an accreton disk, white dwarf and partially 
by the radiation of the red giant. We facilitate the model assuming a
single spherical source of the black body radiation with 
the temperature $T_{\rm s}=30000$~K and luminosity 
$L_{\rm s}=300L_{\odot}$; these parameters are similar to those of the symbiotic 
recurrent nova RS Oph between outbursts (Dobrzycka et al 1996). 
Depending on the 
radial extent of the Str\"{o}mgren zone the wind is fully or partially 
ionized. In order to model 
ionization in the partially ionized wind we adopt two levels plus 
continuum approximation. In this case the ionization can occur 
primarily via the two-stage process:  excitation of the second level 
by the radiation of the central source with the subsequent photoionization 
from the second level. The L$\alpha$ radition transfer is 
treated in the frame of a simple probablity approach, which is
widely used in the modelling of spectra of active galaxies 
(Collin-Souffrin \& Dumont 1989).
The escape probability for L$\alpha$ quanta is adopted to be 
 $\beta_{12}=1/(1+\tau_{12})$, where 
 $\tau_{12}=\tau_{12,\rm in}+\tau_{12,\rm out}$,  
$\tau_{12,\rm in}$ is the optical distance of a given point of the wind 
from the inner wind boundary, while $\tau_{12,\rm out}$ is the 
optical distance from a given point to the outer boundary of the wind.
The excitation rate of the second level is determined by the 
probability $\beta_{12,\rm in}=1/(1+\tau_{12,\rm in})$.

The hydrogen ionization is calculated using a steady-state approximation 
in the inner zone, where stationary approximation is valid, and 
using time-dependent model in the outer zone. Taking into account the 
fact that the photoionization rate for the excited level steeply decreases 
with the radius ($P\propto r^{-4}$) and gets essentially small at large radii 
it would 
be of interest to consider a simple recombination model that 
includes only expansion and recombination. Assuming a constant temperature 
the equation for the electron concentration in the co-moving frame 
reads
\begin{equation}
\frac{dn_{\rm e}}{dr}=-\frac{2n_{\rm e}}{r}-an_{\rm e}^2\,,
\end{equation}
where $a=\alpha/u$ ($\alpha$ is the recombination coefficient).
Solving the equation analytically with the boundary condition 
$x=n_{\rm e}/n=x_0$ at the inner radius $r_0$ and taking account of 
the hydrogen concentration $n=n_0(r_0/r)^2$ we find the 
ionization fraction for $r>r_0$
\begin{equation}
x=x_0\left(\frac{r}{r_0}\right)[(1+\xi)\left(
\frac{r}{r_0}\right)-\xi]^{-1}\,,
\end{equation}
where $\xi=\alpha x_0n_0r_0/u$ is stationary ionization parameter.
For $r\gg r_0$ the ionization fraction asymptotically approaches the constant 
value $x=x_0/(1+\xi)$, i.e. a freeze-out occurs (Zeldovich \& Raizer 2002).
The photoionization from the second level somewhat changes this simple relation.

The calculated ionization fraction in the model with the 
temperature distribution $T=T_{\rm s}W^{1/4}$, 
where $W$ is the dilution factor, is plotted in Fig. 1 
for two values of the density parameter 
$\omega$ (0.1, and 1). In the range 
$r>10^{14}$ cm the ionization fraction does not exceed $10^{-2}$.
The weak dependence $x(r)$ in the outer zone $r>10^{15}$ cm 
reflects the freeze-out effect considered above.
For $\omega=1$ the ionization fraction at large distances attains 
$10^{-4}$. In this case the electron concentration is comparable 
with the contribution of singly ionized metals (Mg, Si, Fe). For the 
rarefied wind $\omega=0.01$ the gas is fully ionized. In this case we 
adopt $x=1$ and $T_{\rm e}=10^4$~K.
Hereafter we take into account only two ions for Na and Ca 
(Na\,I, Na\,II and Ca\,II, Ca\,III respectively). This is justified by 
the high ionization potential of Na\,II (47 eV), which makes the 
role of Na\,III negligible. In the case of Ca the situation is 
opposite: the low ionization potential of Ca\,I (6.1 eV) results in 
the negligibly low concentration of Ca\,I, which does not affect the 
fraction of Ca\,II. 
 In all three cases of the wind density 
the ratio Na\,I/Na turns out to be of the order of 
$\xi_{\rm Na,I}\sim10^{-4}$, 
while the ratio Ca\,II/Ca is close to unity ($\xi_{\rm Ca,2}\approx1$).
The calculated preexplosion ionization fraction of 
hydrogen, sodium and calcium are substantially modified after the 
explosion. 

\subsection{Wind after explosion} 

The wind ionization and heating after SN~Ia explosion can be produced 
by (1) absorption of the ultraviolet (UV) supernova radiation; (2) absorption 
of the X-rays emitted by the reverse shock originated from the deceleration 
of the supernova in the wind; (3) Compton scattering of gamma-rays from 
$^{56}$Ni - $^{56}$Co - $^{56}$Fe radioactive decay.   
The possibiliy of the wind ionization around SN~2006X by X-rays has been 
already invoked for the interpretation of the CS absorption lines 
(Patat et al. 2007), while the role of gamma-rays in the wind ionization 
has never been discussed. 
The hydrogen photoionization is not efficient, because the emergent supernova 
radiation is strongly suppressed in the UV band due to metal line opacity. 
Estimates based on the model UV spectrum of SN~Ia (Pauldrach et al. 
1996) show that even in the light maximum the supernova radiation 
cannot ionize the hydrogen in the wind. However, the 
photoionization of Na\,I and Ca\,II is of cruial importance 
and will be taken into account.

Interaction of a supernova envelope with a wind results in the 
formation of a double shock structure consisting of the forward 
shock in the wind and the reverse shock in the supernova ejecta. For 
the expected wind densities ($\omega\leq1$) X-rays from the reverse shock 
dominate. To calculate the X-ray luminosity we use a self-similar 
solution describing ejecta deceleration in the wind 
(Chevalier 1982; Nadyozhin 1985). 
The density distribution in the undisturbed ejecta is assumed to be 
power law $\rho\propto v^{-k}$ in the outer layers ($v>v_0$) and 
homogeneous inside ($v<v_0$). 
We use $k=9$ which reproduces the initial deceleration, 
computed for the realistic ejecta density distribution. Parameter 
$v_0$ is determined by the supernova kinetic energy and mass which 
are adopted to be $E=1.4\times10^{51}$ erg and $M=1.4~M_{\odot}$ 
respectively. The density in the reverse shock wave is assumed to be 
constant, while the shock width is taken to be $\Delta R=0.03R$ 
(Chevalier 1982). For the relevant conditions Coulomb collisions 
are not able to equilibrate electron and ion temperatures, so we adopt 
rather arbitrary $T_{\rm e}=0.1T_{\rm i}$. Variations of the 
$T_{\rm e}/T_{\rm i}$ ratio do not affect seriously final 
results.

Of the absorbed X-ray energy the fraction $\eta_{\rm h}\approx x^{0.24}$ 
is spent on heating (cf. Kozma \& Fransson 1992). The rest of the 
absorbed energy is distributed equally between hydrogen 
ionization and excitation so that the efficiency of the hydrogen 
ionization is $\eta_{\rm i}=0.5(1-\eta_{\rm h})$. In addition, 
some fraction of the deposited energy is spent on the two-stage 
ionization via the excitation by secondary electrons and subsequent 
photoionization by the supernova radiation. 
The probability of this channel is 
$\phi_2=P_2/(P_2+A_{21}\beta_{12}+A_{2q})$, 
where $P_2$ is the photoionization rate,
$A_{21}\beta_{12}$ is the L$\alpha$ escape rate, and 
$A_{2q}$ is the two-photon decay probability.
Given the deposition rate $\epsilon$ (erg cm$^{-3}$ s$^{-1}$), 
the ionization rate is then 
$\eta_{\rm i}(1+\phi_2)\epsilon/\chi$, where $\chi$ is the 
hydrogen ionization potential. We assume that equilibration 
of electron and ion temperatures in the wind occurs quickly, --- an acceptable 
approximation for the wind density  $>10^4$ cm$^{-3}$. For instance, 
assuming $n\sim10^4$ cm$^{-3}$ and $T_{\rm e}=4\times10^5$~K the 
equilibration time is $t_{\rm eq}\sim3$ days. 
Radiative cooling is described in terms of the cooling function which 
is taken according to Sutherland \& Dopita (1993). Additionally, we 
take into account the Compton cooling due to photons scattering off the 
wind electrons; this mechanism can contribute markedly in the 
gas cooling for $T_{\rm e}\geq10^5$~K. 

Wind ionization and heating by gamma-rays of the radioactive decay 
is calculated in the approximation of a single scattering which 
is treated as an absorption with the effective absorption coefficient 
$k=\alpha k_{\rm T}$ (where $k_{\rm T}$ is the absorption coefficient due to 
Thomson scattering). 
The average value of $\alpha$ in the range of 0.2--2 MeV is about
 0.12 (Sutherland \& Wheeler 1984). With this value of 
$\alpha$ the gamma-ray absorption coefficient is 
$k\approx0.04$ cm$^2$ g$^{-1}$. 
The fraction of the escaping gamma-ray luminosity $q(t)$ was 
calculated for a set of SN~Ia models by H\"{o}flich et al. (1993). 
We approximate this computations by the "average" analytical expression
\begin{equation}
q(t)=0.05\exp[-(t_1/t)^2]+0.95\exp[-(t_2/t)^2]\,,
\end{equation}
where $t_1=20$ d and $t_2=36$ d. The initial mass of radioactive 
$^{56}$Ni is taken to be $0.55~M_{\odot}$ in accordance with 
the empirical bolometric light curves of SN~Ia (cf. Wang et al. 2007).

The ionization fraction and the gas temperature are calculated via 
numerical solution of time-dependent ionization equations 
in the two-level plus continuum approximation and a time-dependent thermal 
balance equation. An example of the radial distribution of the electron 
temperature and hydrogen ionization fraction in the preshock wind 
on day 30 after the supernova explosion is shown in Fig. 2 
for the case of $\omega=0.1$. The valus of $x$ and $T_{\rm e}$ decrease 
with the radius approximately as $r^{-2}$. In the shock wave proximity 
$T_{\rm e}\sim10^5$~K whereas at the distance $10^{17}$ cm the 
temperature drops down to $\sim10^3$~K. Note, the ionization fraction in the 
wind is substantially higher than before the supernova explosion (cf. Fig. 1). 

The relative contribution of the shock X-rays and gamma-rays of 
radioactive decay in the heating and therefore in the 
ionization is demonstrated qualitatively in Fig. 3 which
shows evolution of the temperature in the preshock zone. 
At the early epoch ($t<15$ d) X-rays 
dominate in the heating, while the gamma-rays dominate at the later 
epoch, at least until day 400. The temperature maximum at about 50 d
being apparent for $\omega=0.01$ and $\omega=0.1$ is related to the 
maximum of luminosity of escaping gamma-rays (inset in Fig. 3). 

\subsection{Intensity of Na\,I and Ca\,II lines after supernova outburst} 

The primary ionization mechanism for Na\,I and Ca\,II after the 
SN~Ia explosion is the photoionization from the ground and excited 
states. The specific feature of these ions is the low photoionization 
cross section for the ground level. The role of the two-stage 
process --- the excitation and subsequent photoionization from the excited 
level --- is, therefore, of high imprortance. The latter 
is especially apparent for Na\,I because the UV radiation 
of SN~Ia is strongly suppressed, while the two-stage ionization 
operates via the visible radiation. We assume the 
spectrum of SN~Ia to be black body with the temperature of 10000~K, but 
with the suppressed UV flux. The time-independent, but wavelength-dependent, 
 UV suppression factor is derived from 
the synthetic and observed spectrum of SN~1992A at
the light maximum (Pauldrach et al. 1996). The bolometric 
light curve of SN~Ia is taken according to the observationally recovered 
bolometric luminosity (Wang et al. 2007).
The Na\,I and Ca\,II ionization is considered in the two-level plus 
continuum approximation. The photoionization cross sections for the 
ground level are taken from Verner et al. (1996), while the photoionization 
cross sections for excited levels are assumed to be hydrogenic. 
The ionization fraction of Na\,I and Ca\,II 
is found by numerical solution 
of time-dependent equation of ionization balance on the background of 
the calculated ionization and temperature of the wind.

After the supernova outburst the Na\,I photoionization dominates 
and therefore the Na\,I concentration drops exponentially 
$dn_1/dt\approx-Pn_1$, where $P$ is the 
photoionization rate. When the photoionization from the excited level 
is dominant the ionization rate is $P\sim10^5W^2\sim10$ s$^{-1}$, assuming
the dilution factor $W=0.01$. On the other hand, the recombination rate 
$\alpha n_{\rm e}\sim4\times10^{-8}$ s$^{-1}$ adopting 
$n_{\rm e}=10^5$ cm$^{-3}$, i.e. by the eight orders lower.
This explains why the Na\,I fraction is small and does not exceed 
$\xi_{\rm Na,1}\sim10^{-8}$ on day 30 (Fig. 4). The Ca\,II photoionization 
is less efficient so the Ca\,II fraction 
in the inner zone is $\xi_{\rm Ca,2}\sim10^{-4}-10^{-2}$, while 
in the outer zone $r>10^{17}$ cm approaches unity.

Evolution of the optical depth in resonant Na\,I É Ca\,II lines 
on the background of the expanding supernova is shown in Fig. 5 for 
three values of the wind density parameter. Inset shows the evolution of 
the radius of the supernova-wind interface which is the 
lower limit for the integration of the optical depth in CS lines. 
When computing the line absorption coefficient we
 used the Dopler width $u_{\rm D}=10$ km s$^{-1}$.
The major conclusion from the presented results is the low 
optical depth of the wind in the Na\,I 5890 \AA\ line ($\tau<10^{-3}$) 
and rather high optical depth in the Ca\,II 3934 \AA\ line ($\tau>0.1$). 
For the same wind density the optical depth in Ca\,II line is by four 
orders larger than that in Na\,I. Interestingly, even for the 
rarefied wind ($\omega=0.01$) the optical depth in the 
Ca\,II 3934 \AA\ line at the early epoch $t\leq10$ d exceeds unity.
These results do not depend on the physical conditions in the pre-explosion
 wind since the photoionization and heating 
after the supernova explosion quickly washes out the initial conditions.

Model uncertainties could affect the calculated Na\,I optical depth, 
although corrections are unlikely exceed the factor of two. 
Remarkably, results for Na\,I are not sensitive to the uncertainty of 
UV spectrum of SN~Ia because the quanta responsible for the photoionization 
from the excited level are distributed in the visible range 
$\lambda>3500$ \AA. The latter statement is not valid for Ca\,II 
because the radiation responsible for the ionization from excited level 
comes from UV band ($h\nu\sim 8-9$ eV). Yet, since 
the UV spectrum used in our model is based on the observational and 
synthetic spectrum of SN~Ia, the uncertainty of UV spectrum unlikely 
markedly change the optical depth in the Ca\,II line compared 
to the above results. 

Of special interest is a model with a dense shell on the background of a
wind of a moderate density $\omega=0.1$. In the considered example 
the shell with the width $\Delta r=0.1r$ is placed at the 
distance $r=6\times10^{16}$ cm. The density of the shell is 
chosen to make the optical depth of Na\,I 5890 \AA\ close 
to unity on day 30. It was found that this condition requires 
the shell mass of $\approx0.5~M_{\odot}$. The evolution of the optical 
depth in Na\,I and Ca\,II lines and the fractions of 
Na\,I and Ca\,II are shown in Fig. 6. This model could relate to 
the interpretation of the CS lines of Na\,I and Ca\,II in SN~2006X.

\section{DISCUSSSION AND CONCLUSION} 

The aim of the paper has been to calculate the intensity of CS 
resonant absorption lines of Na\,I and Ca\,II in the spectrum 
of SN~Ia in the frame of symbiotic scenario. The modelling shows 
that for the wind densities typical of a red giant, 
$\omega\sim 0.01-1$, the expected optical depth in the Na\,I 5890 \AA\ 
line is very low, $\tau<10^{-3}$. The slow wind around symbiotic binary, 
therefore, unlikely can be detected in the absorption Na\,I lines on the 
background of SN~Ia. On the other hand, the model predicts for the same 
CS densities rather strong Ca\,II 3934 \AA\ absorption line with 
$\tau>0.1$. At the early time $t\leq10$ d the optical depth in the 
Ca\,II 3934 \AA\ line can exceed unity even in the case of rarefied wind 
$\omega=0.01$. It is reasonable, therefore, to search for the 
signatures of the red giant wind in the Ca\,II 3934, 3968 \AA\ lines close 
to the light maximum.

The modelling of the circumstellar Na\,I and Ca\,II lines 
in the red giant wind are of particular interest in relation to 
the observations of these lines in SN~2006X. 
In the spectra of this supernova four low velocity variable components 
(A, B, C, D) in Na\,I lines has been found (Patat et al. 2007).
Two (A and B) are observed simultaneously in both Na\,I lines and 
Ca\,II 3934 \AA\ line, while C and D components are seen only in Na\,I lines.
Remarkably, in the spectrum of SN~2006X one does not detect any
absorption components of Ca\,II with the depth $\geq0.2$, which 
is not seen in Na\,I lines (cf. Patat et al. 2007). 
This fact, according to our modelling ( Fig. 5), indicates that 
the wind density around SN~2006X is low, $\omega<0.01$. 
The corresposponing upper limit on the mass loss rate is
$\dot{M}<10^{-8}u_{10}~M_{\odot}$ yr$^{-1}$.

A question arises as to the oigin of the Na\,I and Ca\,II
absorption lines detected in SN~2006X. Components C and D, observed in 
Na\,I lines strengthen between days -2 and +14 relative 
the light maximum, or in the range $16<t<32$ d after the explosion 
(assuming maximum at $t=18$ d) and disappear on day 79 (Patat et al. 2007). 
The disappearence of C and D components might indicate that they 
form in a relatively close $r\sim10^{16}$ cm dense shell that 
is accelerated between days 32 and 79 by the supernova ejecta 
(Patat et al. 2007). A problem with this sceanrio is the absence 
of absorption components C and D in the Ca\,II line. 
Taking into account that in our model the Na\,I line is by 3--4 
orders of magnitude is weaker then the Ca\,II line one needs to admit 
that the latter is strongly suppressed somehow by 4--5 orders of magnitude. 
A reasonable possibility might be depletion of Ca onto dust grains. 
In fact, in the interstellar medium the Ca depletion factor can be as low 
as $10^{-3}-10^{-4}$ (Krinklow et al. 1994). However, in the case 
of C and D components this expalnation faces a serious difficulty.  
The point is that even the heavy depletion of Ca were the case, the 
dust would evaporate after the supernova explosion.
Indeed, assuming the grain absorption efficiency 
$Q_{\rm a}\propto 1/\lambda$, supernova luminosity $10^{43}$ erg s$^{-1}$, 
and the supernova radiation tempearture $T=10^4$~K one gets the 
dust temeperature $T_{\rm d}=W^{1/5}T\approx2600$~K at the 
radius $1.5\times10^{16}$ cm. The derived dust temperature is 
substantially higher than the evaporation temperature ($\sim1000$~K).
This implies that the shell with the radius of $\sim10^{16}$ seems to 
be not viable model for the C and D components.

The second pair of components, A and B, which are seen in both 
Na\,I and Ca\,II is presumably related with the distant shell 
(Patat et al. 2007), because on day 139 the lines retain the same 
intensity as at the previus epoch. For the wind density 
parameter $\omega=0.1$ the radius of SN-wind interface suggests 
that the shell radius should be $r>4\times10^{16}$ cm 
(cf. Fig. 5). The model of the Na\,I and Ca\,II line formation in the 
shell with the radius of $6\times10^{16}$ cm illustrates this 
possibility. However in this case the shell mass should be rather large,
$\approx0.5~M_{\odot}$. The origin of this shell is a not a trivial problem.
As a possibility one could speculate that the shell has been ejected 
as a result of a robust dynamical mass loss episode.

Difficulties of the interpretation of Na\,I and Ca\,II absorption 
lines in the SN~2006X spectrum in the model of the CS wind 
push us to an alternative view, which suggests that these lines 
originate beyond the dust evaporation radius $r>10^{17}$ cm.
This conjecture permits us to explain why Ca\,II lines are not seen in 
components C and D. In this case  the variability of the Na\,I line 
depth is presumably caused by the variation of occultation by clouds
of the expanding supernova "photosphere". The model of 
variable covering factor due to the photosphere expansion was already 
discussed and rejected by Patat et al. (2007) because A and B components 
of Ca\,II line did not show simultaneous variability. However, the latter 
argument loses its strength, if there are clouds of two types: 
(1) "ordinary" clouds in which Ca is not depleted, 
with the weak Na\,I and strong Ca\,II lines;
(2) dense dusty clouds in which Na\,I lines are strong but  
Ca\,II lines are suppressed owing to Ca depletion onto dust grains. 
In this case the behavior of Ca\,II and Na\,I lines can be independent 
with the Ca\,II lines being constant, and Na\,I being variable; the latter, 
e.g., because the early supernova photosphere is not covered by 
the second sort of clouds. 

Components C and D which are seen only in Na\,I line could be explained 
in a similar way; in this case however only clouds of second variety are 
needed. Since C and D components do not show Ca\,II lines these clouds 
should be dusty in order to deplete Ca and be at distances larger than 
the dust evaporation radius. The evolution of components C and D in 
Na\,I lines could be expalined in the following way. At the beginning,  
when the photosphere radius is small, the supernova is seen presumably 
through the hole of some cluster of clouds. The supernova photosphere 
while expanding gets occulted by clouds so the covering factor increases. 
Assuming that the cloud cluster occupies a finite solid angle in the sky plane 
one comes to the picture when a subsequent expansion of the photosphere 
would result in the covering factor decrease. The performed 
modelling shows that the required evolution of components C and D can 
be reproduced, e.g., by the cluster of four clouds with a typical 
cloud radius of $\sim1.3\times10^{15}$ cm. 

The origin of absorbing clouds at the radii 
$r>10^{17}$ cm is unclear. They may originate from the matter lost by 
the binary system in which SN~2006X exploded. Alternatively they 
may not be related to the presupernova of SN~2006X at all.
In this regard it is noteworthy that SN~2006X shows unusually high 
reddening $E(B-V)\approx1.4$ (Wang et al. 2007). The presence of 
large amount of clouds in the line of sight might create a favourable 
condition for the variability of Na\,I and Ca\,II lines. 
If so, one could expect the similar variability of Na\,I and Ca\,II lines
 in other cases of SN~Ia with heavy reddening. 
It should be emphasised that clouds responsible for the variation 
of Na\,I and Ca\,II lines provide only small fraction of the total 
dust absorption. This follows from the equivalent width of 
A, B, C, and D components of Na\,I line, 
 which is a small fraction of total equivalent 
width of the Na\,I absorption 
line in our and host galaxies (Patat et al. 2007).  

Summing up, a signature of a red giant wind around 
SN~Ia exploded in a symbiotic binary will be unlikely 
observable in the Na\,I absorption lines, while the Ca\,II absorption 
lines can be detectable. The absence in the 
SN~2006X spectrum of the Ca\,II absorption lines which are not related with 
the similar componets of Na\,I lines implies the upper limit 
 on the mass loss rate 
 $\dot{M}<10^{-8}u_{10}~M_{\odot}$ yr$^{-1}$. The absorption lines 
 of  Na\,I and Ca\,II observed in SN~2006X are related probably 
 with distant clouds at the radii $r>10^{17}$ cm. 

\section{Acknowledgements}
I am grateful to Nando Patat for discussions.

{}


\clearpage

\begin{figure}
\plotone{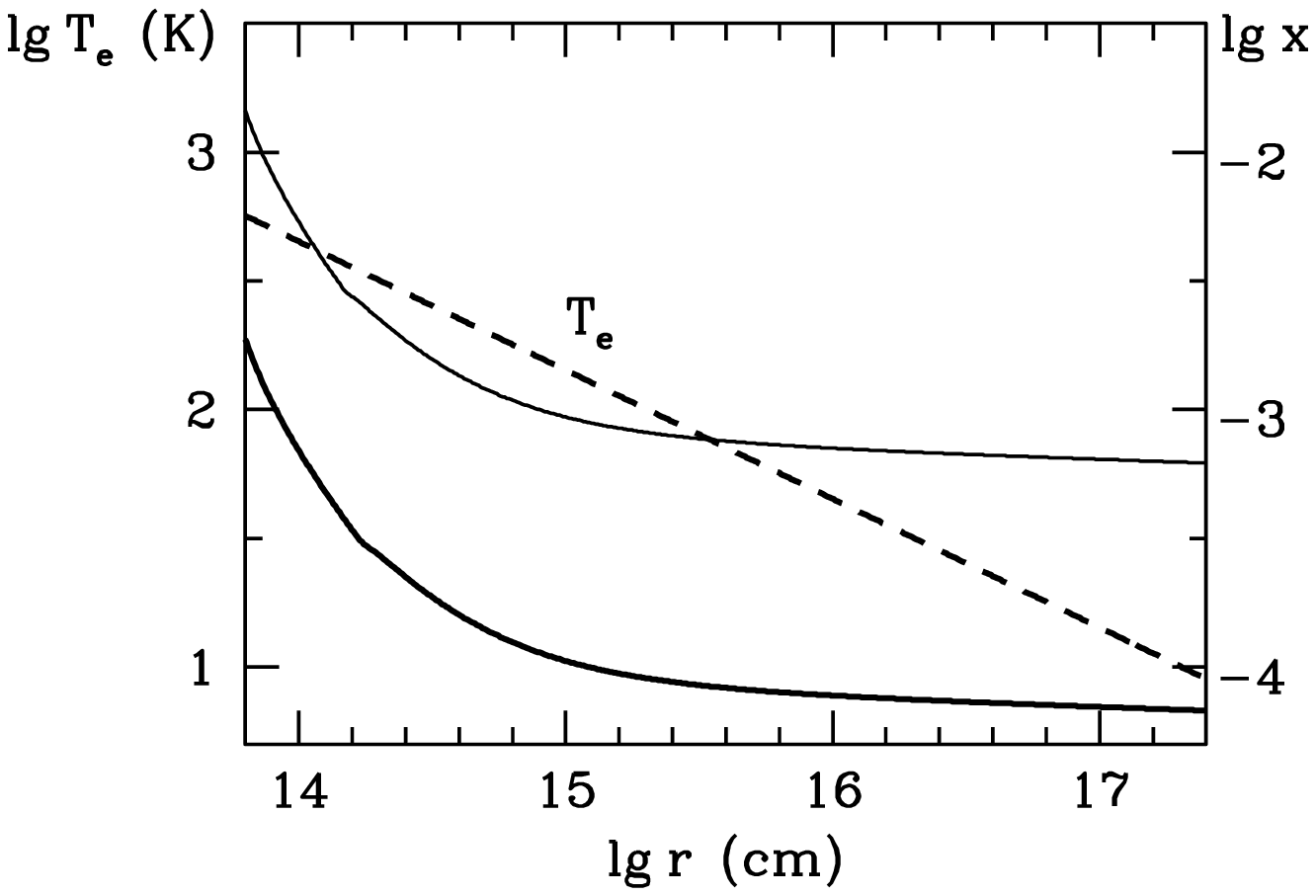}
\caption{
Ionization fraction and temperature in the wind prior to  
supernova explosion for two values of the wind density parameter
$\omega=0.01$ and $\omega=0.1$. The thinnest line refers to lower density
}
 \end{figure}

\begin{figure}
\plotone{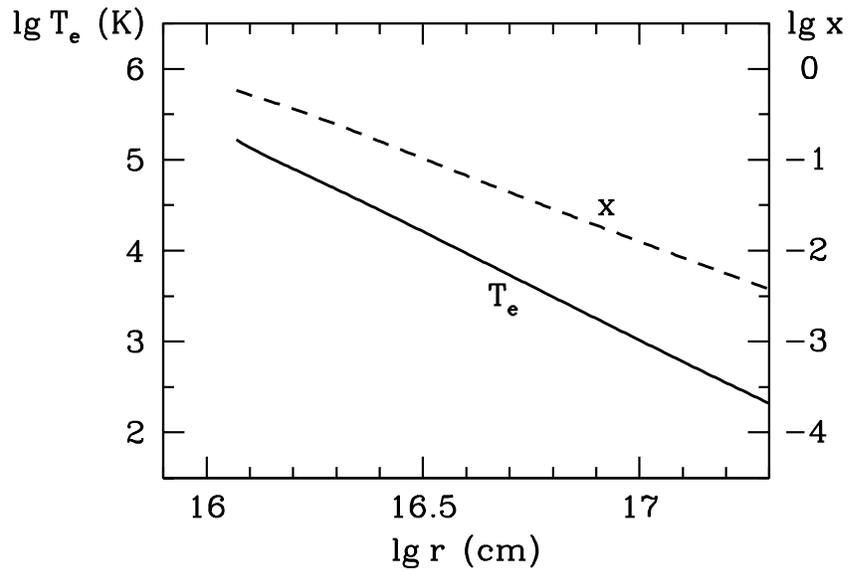}
\caption{
Ionization fraction and temperature in the wind on day 30 after the 
supernova explosion for the wind density parameter $\omega=0.1$.
}
 \end{figure}

\begin{figure}
\plotone{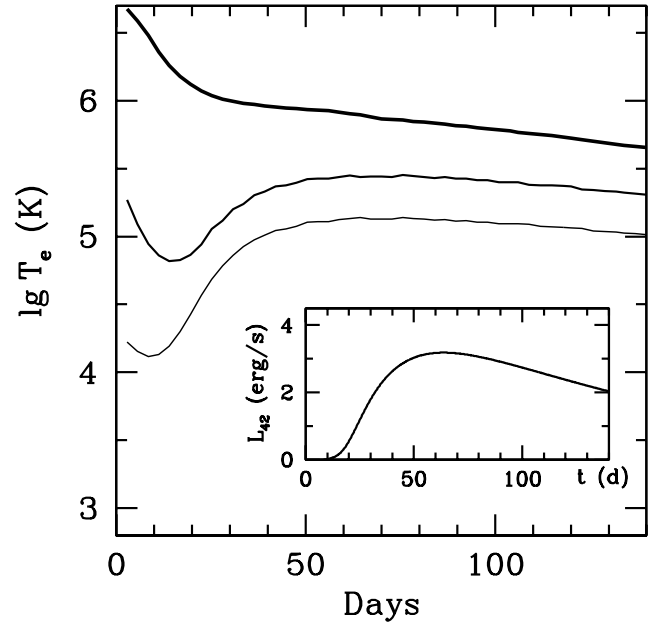}
\caption{
Preshock temperature in the wind for three values of the wind density 
parameter  $\omega=0.01$, 0.1, and 1. The line thickness grows with 
the density. The inset shows the evolution of the luminosity of 
escaping gamma-rays in units of $10^{42}$ erg s$^{-1}$.
}
 \end{figure}

\begin{figure}
\plotone{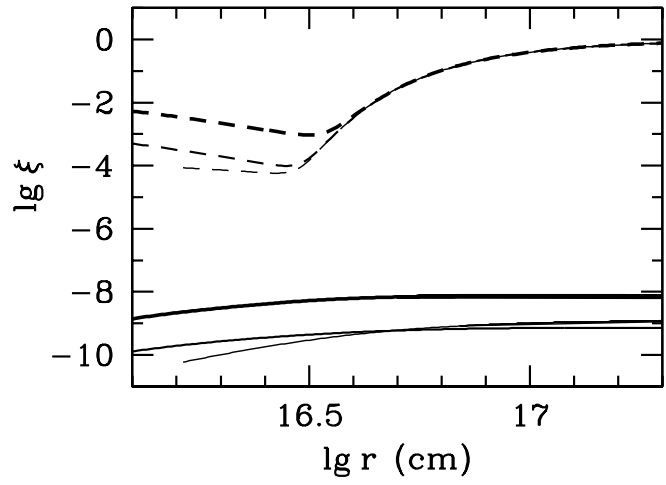}
\caption{
The ionization fraction of Na\,I ({\em solid line}) and Ca\,II on day 30 
after the supernova explosion for three values of the wind density 
parameter $\omega=0.01$, 0.1, and 1. The line thickness grows with 
the density.
}
 \end{figure}

\begin{figure}
\plotone{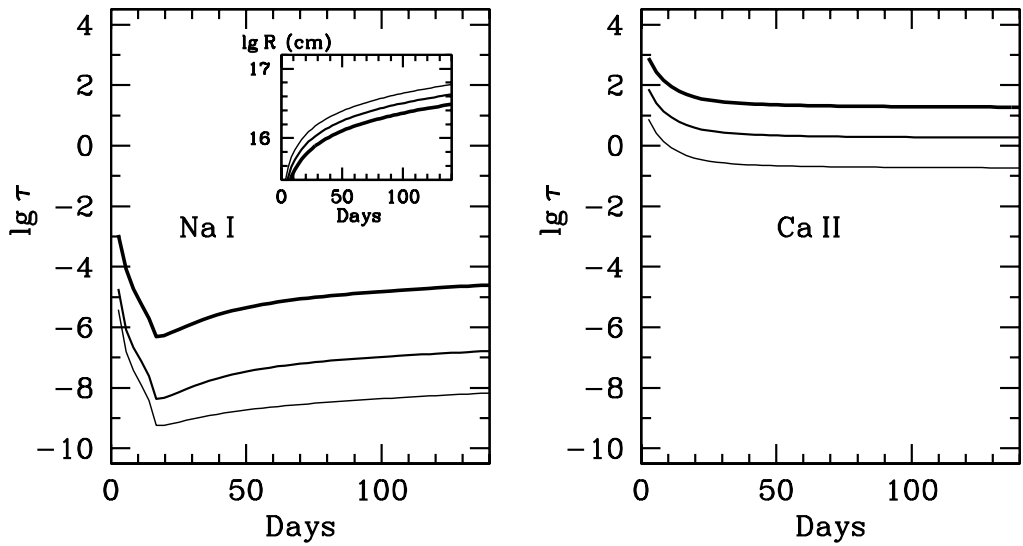}
\caption{
The model evolution of the optical dept in the Na\,I 5890 \AA\ 
(left) and Ca\,II 3934 \AA\ (right) for three values of the 
wind density parameter $\omega=0.01$, 0.1, and 1.
The line thickness grows with the density.
The inset shows the evolution of the radius of the SN-wind interface
for the wind densities indicated above. 
}
 \end{figure}

\begin{figure}
\plotone{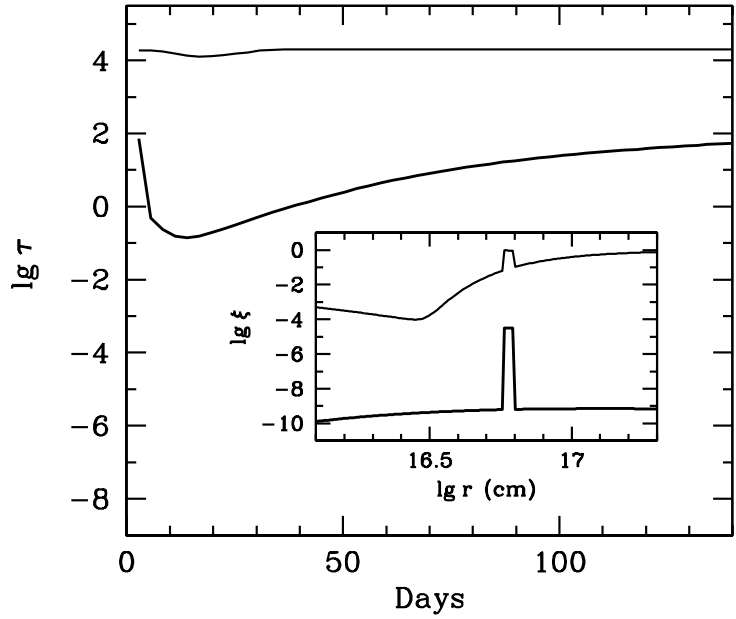}
\caption{
The model evolution of the optical depth in the Na\,I 5890 \AA\ 
 ({\em thick line}) and Ca\,II 3934 \AA\ for the 
 wind density parameter $\omega=0.1$ in the case of the dense shell 
 at the distance $r=6\times10^{16}$ cm. 
 The inset shows the radial dependence of the fraction of Na\,I 
 ({\em thick line}) and Ca\,II on day 30 after the supernova explosion.
}
 \end{figure}

\end{document}